# Different Approaches to Community Evolution Prediction in Blogosphere

Bogdan Gliwa[1], Piotr Bródka[2], Anna Zygmunt[1], Stanisław Saganowski[2], Przemysław Kazienko[2], Jarosław Koźlak[1]

[1] AGH University of Science and Technology, Al. Mickiewicza 30, 30-059 Kraków, Poland

[2] Institute of Informatics, Wrocław University of Technology, Wyb.Wyspiańskiego 27, 50-370 Wrocław, Poland

bgliwa@agh.edu.pl, piotr.brodka@pwr.wroc.pl, azygmunt@agh.edu.pl, stanislaw.saganowski@pwr.wroc.pl, kazienko@pwr.wroc.pl, kozlak@agh.edu.pl

*Abstract* **Predicting the future direction of community evolution is a problem with high theoretical and practical significance. It allows to determine which characteristics describing communities have importance from the point of view of their future behaviour. Knowledge about the probable future career of the community aids in the decision concerning investing in contact with members of a given community and carrying out actions to achieve a key position in it. It also allows to determine effective ways of forming opinions or to protect group participants against such activities. In the paper, a new approach to group identification and prediction of future events is presented together with the comparison to existing method. Performed experiments prove a high quality of prediction results. Comparison to previous studies shows that using many measures to describe the group profile, and in consequence as a classifier input, can improve predictions.**

## I. Introduction and Related Work

In recent years dozens of community extraction methods have been developed, also several methods to track changes of the group over the time have been presented. Lately, one of the most investigated aspect of social network analysis is prediction. The best investigated is link prediction problem [1] [2] [3]. It refers to predicting the existence of a link (relation) between two nodes (users) within a social network. Prediction is being made based on different network and group measures. For example Liben-Nowell et al. [1] focused on path and common neighbours between pair of nodes, while Lichtenwalter et al. [3] consider degrees and mutual information between them. Zheleva et al. [2] explored different combinations of descriptive, structural and group features (e.g. group membership) and proved that prediction accuracy is 15% - 30% more accurate as compared to using descriptive node attributes and structural features.

After successful results in link prediction the researchers have immersed the problem to link sign prediction [4] [5] [6] [7]. Sign in this context means that predicted relation between users may be positive or negative. Again the prediction is being made based on network and group measures. Symeonidis et al. [4] look at paths between the node pair and use the notion of similarity to predict the sign. Leskovec et al. [5] use degree and mutual information between pair of nodes for link prediction and profits from the theory of balance and status to predict the link sign. Kunegis et al. [7] evaluated different signed spectral similarity measures to predict the sign of the link in Slashdot.

Davis et al. [8] tackled the problem of multi-relational link prediction by extending the neighbourhood methods with weight and focusing on triads. Richter et al. [9] and Wai-Ho et al. [10] faced the very important task of churn prediction. Wai-Ho et al. introduced a new data mining algorithm called DMEL (data mining by evolutionary learning), which estimates each prediction being made. Richter et al. presented a novel approach and tried to predict churn based on analysis of group behaviour. This approach touches another aspect, not well studied yet, where evolution of the whole group is being predicted, i.e. which event will be next in group lifetime.

Despite a lot of time spent on literature review, the authors could not find any methods referring to the group evolution prediction problem (except the method presented in [17]). Therefore this article focuses precisely on the prediction of the group evolution. A new method for future event prediction based on stable group changes identification algorithm (*SGCI*) has been developed. Prediction in this method is being made based on the previous events in group lifetime extracted by *SGCI* and group profile described by group size, cohesion, leadership and density. Additionally the comparison to the previous method described in [17] were performed. The results shows that using many measures to describe the group profile, and in consequence as a classifier input, can improve prediction.

## II. Methods of Events Identification in Group Evolution

### A. SGCI: The algorithm for stable group changes identification

The algorithm for the identification of states of the groups consists of the following steps:

**Step 1.** Identification of fugitive groups in the separate time frames.

Whole network is divided into time frames and in each time frame the method of finding communities in network is applied.

**Step 2.** Identification of group continuation – assigning transitions between groups in neighbouring time steps.

The work was partially supported by The National Science Centre - the research project, 2010-2013, Institute of Informatics of Wrocław University of Technology, and The European Commission under the 7th Framework Programme, Coordination and Support Action, Grant Agreement Number 316097, ENGINE - European research centre of Network intelliGence for INnovation Enhancement http://engine.pwr.wroc.pl/.

After extracting communities in time frames, the communities from neighbouring time frames are matched and algorithm assigns transitions between them (from group in time frame $t$ to group in time frame $t+1$). It is carried out by calculating the Modified Jaccard Measure (A and B are groups and |A| means size of group A, A and B are not empty) for each pair of groups from neighbouring time slots:

$$MJ(A,B) = \max\left(\frac{|A \cap B|}{|A|}, \frac{|A \cap B|}{|B|}\right)$$

and if the value of this measure is above a defined threshold (for the tests we assumed value 0.5) and difference in size between these groups

$$ds(A,B) = \max\left(\frac{|A|}{|B|}, \frac{|B|}{|A|}\right)$$

is no more than specified value (in experiments no more than 50 times one group is bigger than the other one), then the algorithm make transition between these groups.

**Step 3.** Separation of the stable groups (lasting for at least required subsequent time steps).

In this step, the stable groups are retrieved. It is conducted by rejecting groups which do not exist in the required number of subsequent time frames (in the experiments we required that a group should exist in at least three subsequent time frames).

**Step 4.** Identification of types of group changes. Assigning events from the list describing the change of the state of the group to the transitions

Each transition between stable groups from neighbouring time frames is associated with the attribute describing the kind of change of the group.

We can define some types of group changes (in the following notation we assume for a given transition that A means group from the first time frame in this transition and B means group from the second time frame in this transition, sh and dh are some thresholds; for experiments sh=10 and dh=0.05):

1. **addition** - takes place when a small group attaches to a large one:

$$\frac{|B|}{|A|} \geq sh$$

2. **deletion** - when a small group detaches from a large one:

$$\frac{|A|}{|B|} \geq sh$$

3. **merge** - many groups in one time frame form a new larger group in the next time frame. Transition is one among few transitions such ones that difference in size between groups from different time frames is no more than $sh$ times and first group in transition is smaller than the second one:

$$|A| < |B| \wedge ds(A,B) < sh$$

4. **split** – group divides into some smaller groups in next time frame. Transition is one among few such ones that difference in size between groups from different time frames is no more than sh times and first group in transition is larger than the second one:

$$|A| > |B| \wedge ds(A,B) < sh$$

5. **split_merge** - occurs when a group divides into at least 2 groups in the next time frame and one of this groups from next time frame is a result of merging with another from a previous time frame

6. **constancy** - simple transition without significant change of the group size:

$$\frac{abs(|A|-|B|)}{|A|} \leq dh$$

7. **change size** – simple transition with the change of the group size :

$$\frac{abs(|A|-|B|)}{|A|} > dh$$

8. **decay** - group does not exist in next time frame.

For a given group it is possible to match more than one event from this group to groups in the next time frame. Some events can coexist with other ones but some of them cannot. If group has assigned constancy event, then there cannot be assigned change size (what is obvious), merge or split event, but there can be addition or deletion events. The analogous case is for co-occurrence events with the change size event. Generally, the addition and the deletion events can coexist with each event type, except the decay event. The decay event is always a single event for the group.

*SGCI* method was introduced in [11], but its predecessor was method described in [12]. In [13] authors described a tool for visualisation of group evolution based on *SGCI* method.

*B. GED: Group Evolution Discovery*

The *GED* method utilize a measure called *inclusion*. It allows to evaluate the inclusion of one group in another. The inclusion of group $G_1$ in group $G_2$ is calculated as follows:

$$I(G_1,G_2) = \underbrace{\frac{|G_1 \cap G_2|}{|G_1|}}_{\text{group quantity}} \cdot \underbrace{\frac{\sum_{x \in (G_1 \cap G_2)} NI_{G_1}(x)}{\sum_{x \in (G_1)} NI_{G_1}(x)}}_{\text{group quality}}$$

where $NI_{G_1}(x)$ is the value reflecting importance of the node $x$ in group $G_1$.

The *GED* method takes into account both the quantity and quality of the group members. The quantity is reflected by the first part of the inclusion measure, whereas the quality is expressed by the second part of the inclusion measure and can be expressed by any user importance measure e.g. centrality degree, betweenness degree, page rank, social position etc.

It is assumed that only one event may occur between two groups ($G_1$, $G_2$) in the consecutive time frames, however one group in time frame $T_i$ may have several events with different groups in $T_{i+1}$. The event type is assigned based on the value of the inclusion measures and size of the groups. Possible events are: continuing, shrinking, growing, splitting, merging, dissolving, forming. The detailed explanation of *GED* method can be found in [14].

### C. Event Types Comparison

In [11] we compared these 2 methods of identification events in group evolution. For events from *GED* we can match the corresponding (to some extent) events from *SGCI*, what is summarized in Table 1.

TABLE 1. CORRESPONDING EVENTS IN DIFFERENT METHODS

| *GED* event | *SGCI* event |
|---|---|
| **continuing** | constancy |
| **growing+shrinking** | change size |
| **merging** | merge+addition |
| **splitting** | split+deletion |
| **dissolving** | decay |

Some events cannot be matched e.g. split_merge from *SGCI*.

### III. PREDICTING GROUP EVOLUTION IN THE SOCIAL NETWORK

#### A. Predicting Group Evolution Using SGCI Results

Presented approach was used in conjunction with *SGCI* method for identification of groups events, but the approach can be used with any method of identification of changes in group evolution. This approach for prediction future events of groups employs classifier.

The approach is based on sequences of 3 states of groups (one state from present time frame for given group and two states from preceding time frames for predecessors of given group). Figure 1 explains notion of sequences and input data for classifier. In this figure 3 sequences are marked (labelled as *seq1*, *seq2* and *seq3*). For instance, *seq1* is a sequence of states of groups $G_{n-2,1}$, $G_{n-1,1}$ and $G_{n,1}$. The state of each group is described by the following measures:

- **leadership** - measure describing centralization in graph or group (the largest value is for star network) [15]

$$L = \sum_{i=1}^{n} \frac{d_{max} - d_i}{(n-2)(n-1)}$$

where $d_{max}$ means maximum value of degree in group, $n$ - number of nodes in group,

- **density** - measure expressing how many connections between nodes are present in network in relation to all possible connections between them [16]

$$D = \frac{\sum_i \sum_j a(i,j)}{n(n-1)}$$

where function $a(i,j)$ has value 1 when there is connection from node $a$ to node $b$,

- **cohesion** - measure characterising strength of connections inside group in relation to connections outside group (from group members) [16]

$$C = \frac{\frac{\sum_{i \in G} \sum_{j \in G} w(i,j)}{n(n-1)}}{\frac{\sum_{i \in G} \sum_{j \notin G} w(i,j)}{N(N-n)}}$$

where $w$ is function assigning weight between nodes, $G$ is group, $n$ - number of nodes in group and $N$ - number of nodes in network,

- **group size** - number of nodes in group.

Described sequence of group states is an input for classifier. The predicted variable is the dominating next event for the last group in a sequence. For instance, for sequence *seq1* we want to predict the next evolution event for group $G_{n,1}$. As we can see in Figure 1, this group has 2 events assigned: change size (transition between $G_{n,1}$ and $G_{n+1,1}$) and addition (transition between $G_{n,1}$ and $G_{n+1,2}$). That is the reason we introduced the concept of dominating event (we can only predict one event per group).

The dominating event is one of events assigned for a given group. Such an event is determined on the basis of priority of existing events - the event with the highest priority is chosen. We use the following order of events (from the highest priority to the lowest one): constancy, change size, split, merge, addition, deletion, split_merge, decay. The reason for this order is that some events such as addition or deletion mean small change for groups. Moreover, some events cannot coexist with other ones (described in section IIA) and position in order of such events is meaningless (such as the decay event).

For instance, the group $G_{n,1}$ has 2 assigned events: change size and addition, so the dominating event for group $G_{n,1}$ is change size because this event has higher priority.

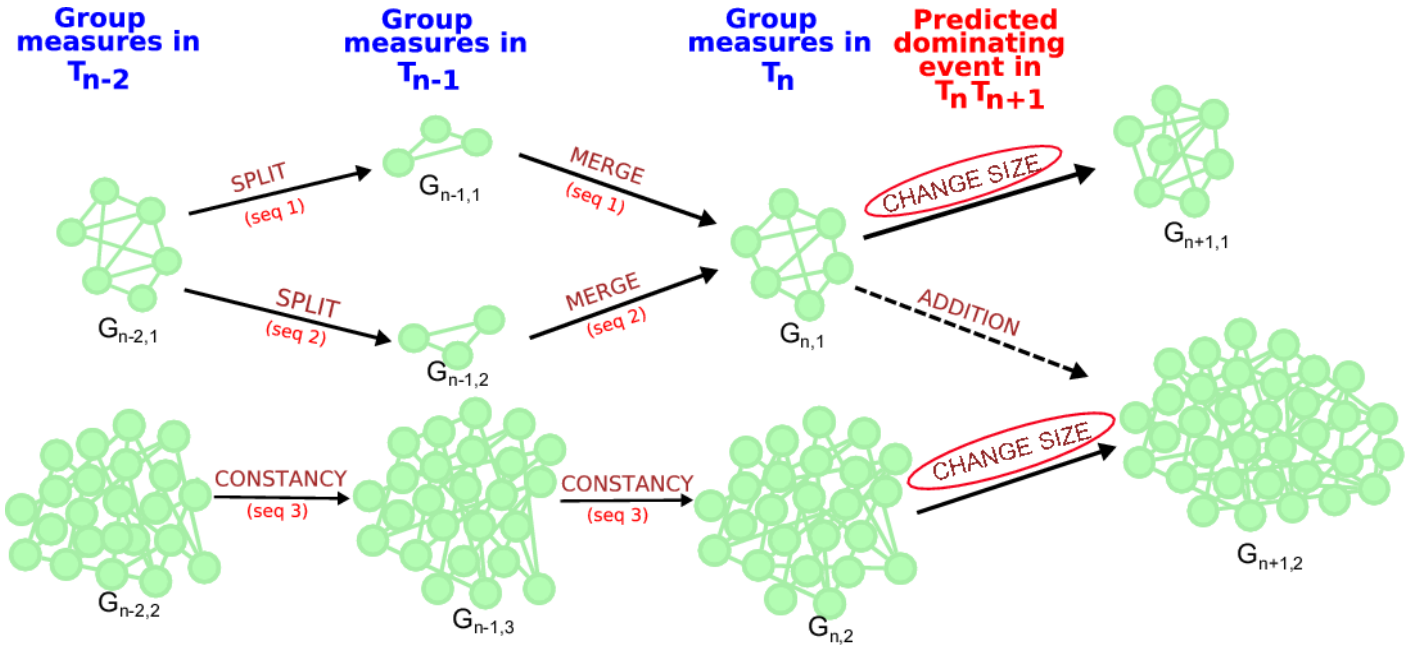

**Figure 1.** Explanation of input data of classifier for prediction purpose - sequences of group measures from 3 time frames (1 present group state and 2 earlier group states) and predicted dominating event.

*B. Predicting Group Evolution Using GED Results.*

This approach was presented in [17] and involves usage of the results of *GED* method [14]. The idea is that using a simple sequence, which consists of several preceding groups profiles and events, as an input for the classifier, the learnt model will be able to produce very good results even for simple classifiers.

The sequences of groups sizes and events between time frames can be extracted from the *GED* results. In this paper 4-step sequences are used (Figure 2). Obviously, the event types vary depending on the individual groups, but the time frame numbers were fixed. It means that for each event four group profiles in four previous time frames together with three associated events are identified as the input for the classification model, separately for each group. A single group in a given time frame ($T_n$) is a case (instance) for classification, for which its event $T_nT_{n+1}$ is being predicted.

The sequence presented in Figure 2 is used as an input for classification. The first part of the sequence is used as input features (variables), i.e. (1) **Group profile in $T_{n-3}$**, (2) **Event type $T_{n-3}T_{n-2}$**, (3) **Group profile in $T_{n-2}$**, (4) **Event type $T_{n-2}T_{n-1}$**, (5) **Group profile in $T_{n-1}$**, (6) **Event type $T_{n-1}T_n$**, (7) **Group profile in $T_n$**. A predictive variable is the next event for a given group. Thus, the goal of classification is to predict (classify) **Event $T_nT_{n+1}$ type** – out of the six possible classes: i.e. (1) growing, (2) continuing, (3) shrinking, (4) dissolving, (5) merging and (6) splitting. Forming was excluded since it can only start the sequence.

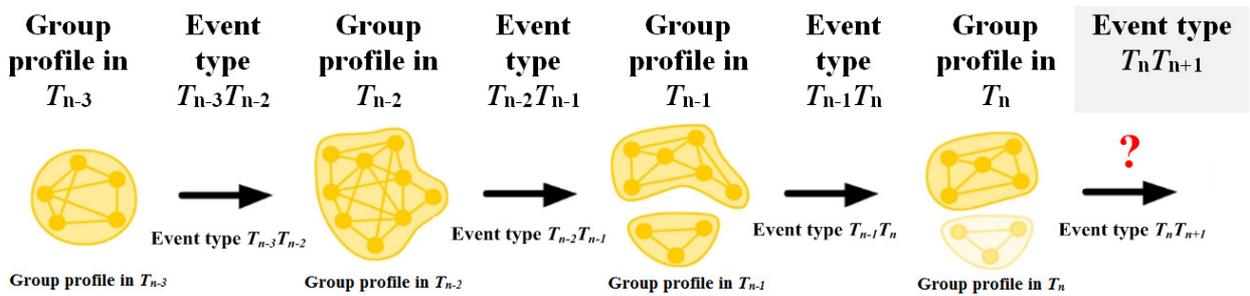

**Figure 2.** The sequence of events for a single group together with its profiles as well as its target class - event type in $T_nT_{n+1}$. It corresponds to one case in classification

## IV. DATASET AND EXPERIMENT SETUP

*A. Dataset description*

Experiments were conducted on data from the portal www.salon24.pl which contains many blogs, most of them are political blogs. The data consists of 26 722 users, 285 532 posts and 4 173 457 comments. For tests we used data from range 4.04.2010 - 31.03.2012 (half of data set). The analysed period of time was divided into time frames, each lasting 7 days and neighbouring time frames overlap each other by 4 days. In this period of time there are 182 time frames.

## B. Group extraction

After separation of time frames the groups were extracted in each of the time frames. We used CPM method (CPMd version) from CFinder tool (http://www.cfinder.org/) for k=5.

## C. Group sizes

As we can notice in Figure 3 there are many small groups and groups with size 5 outnumber other ones.

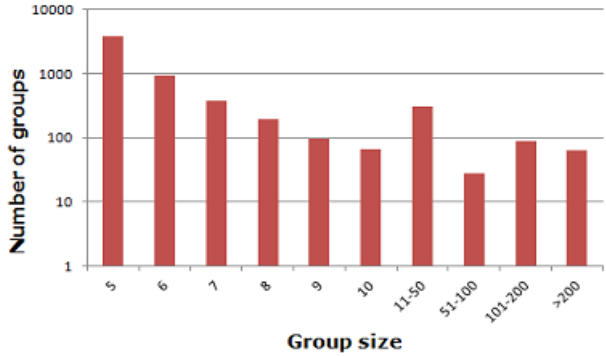

**Figure 3.** Number of groups with given size

## D. Experiment setup

The experiments using *SGCI* method were conducted using following parameters (described earlier in IIIa): *MJ*=0.5, *ds*=50, *sh*=10 and *dh*=0.05.

The *GED* method was run on the dataset with all combination of *GED* parameters [14] from the set {50%, 60%, 70%, 80%, 90%, 100%}. As a node importance measure the social position measure [18] (measure similar to page rank) was utilized.

To describe the group profile, its size, density, cohesion and leadership were used.

The experiment was executed in KNIME (www.knime.org) with Weka plugin. Seven different classifiers were utilized (Table 2) with default settings. For the method of validation 10-fold cross-validation was used [26] with stratified sampling as a method of sampling from *GED* and *SGCI* results. The measure selected for presentation and analysis of the results is F measure which is the harmonic mean of precision and recall.

TABLE 2. CLASSIFIERS USED

| Short Name | Name |
|---|---|
| CART | Classification And Regression Tree [19] |
| J48 | C4.5 decision tree [20] |
| RandomForest | Random forest [21] |
| BayesNet | Bayes network classifier [22] |
| NaiveBayes | Naive Bayesian classifier [23] |
| Ibk | k-nearest neighbour classifier [24] |
| DecisionTable | Decision table (rule classifier) [25] |

## V. EXPERIMENTS

All classifiers were utilized for both approaches, the results are presented below.

### A. Predicting Group Evolution Using SGCI Results

Table 4 presents results of prediction events for different classifiers. Tree classifiers (J48, Random Forest and Simple CART) and Decision Table (rule classifier) achieved the best results. Notably worse results are for Naive Bayes and IBk.

TABLE 4. F-MEASURE FOR EACH EVENT TYPE (CLASS) AND EACH CLASSIFIER

| Classifier / Event type | J48 | Random Forest | Simple Cart | BayesNet | NaiveBayes | Decision Table | Ibk |
|---|---|---|---|---|---|---|---|
| addition | 0.99 | 0.99 | 0.99 | 0.95 | 0.84 | 0.99 | 0.88 |
| change size | 0.97 | 0.97 | 0.98 | 0.86 | 0.31 | 0.99 | 0.72 |
| constancy | 0.97 | 0.96 | 0.97 | 0.87 | 0.00 | 0.99 | 0.54 |
| merge | 0.95 | 0.95 | 0.95 | 0.81 | 0.08 | 0.99 | 0.57 |
| split | 0.96 | 0.98 | 0.97 | 0.89 | 0.36 | 0.99 | 0.72 |
| deletion | 1.00 | 1.00 | 1.00 | 0.77 | 0.31 | 1.00 | 0.71 |
| decay | 0.71 | 0.77 | 0.69 | 0.51 | 0.38 | 0.63 | 0.63 |

Figure 4 shows results of classification for 3 tree classifiers. One can see that results for these 3 classifiers are very similar - the biggest difference is for the decay event which seemed harder to classify. Other events are well classified.

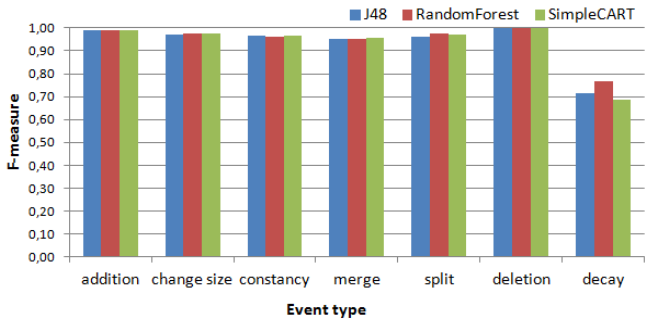

**Figure 4.** Results of event classification for decision tree classifiers

In figure 5 there are displayed results of prediction obtained by probabilistic classifiers. The results vary a lot. BayesNet achieved quite good results, but results of NaiveBayes are much worse. The explanation for bad results of NaiveBayes is that this classifier is based on assumption of independence features used to classification task. This requirement is not met because some values of one measure are correlated with values of another measure e.g. generally density has higher values for smaller groups.

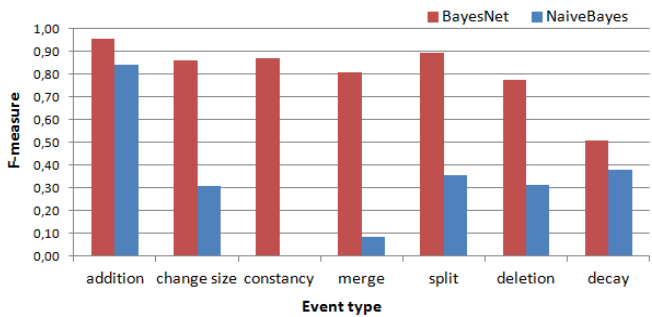

**Figure 5.** Results of event classification for probabilistic classifiers

In figure 6 we can see results for other tested classifiers. Results of the DecisionTable classifier are comparable with results of tree classifiers and, similarly, the decay event is significantly worse classified than other events. The IBk classifier accomplished worse results of prediction than DecisionTable one. For this classifier the hardest event to classify seemed to be constancy.

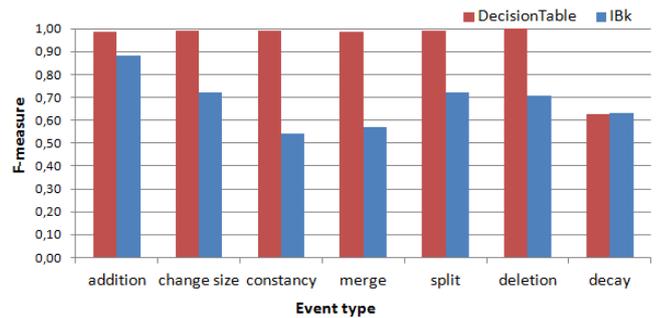

**Figure 6.** Results of event classification for other classifiers

Figure 7 shows distribution of classified events. We can observe that the most popular event is the addition event and there is significantly more events of this type than other types of events. This explains why this event is very well classified for each tested classifier.

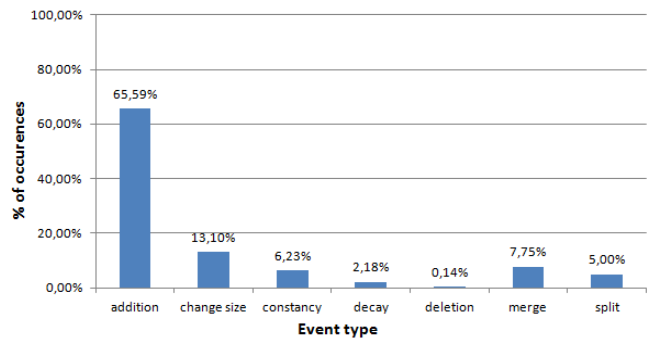

**Figure 7.** The percentage of events in dataset.

### B. Predicting Group Evolution Using GED Results.

Based on the suggestions from [17] the results for *GED* parameters equal to 70 where utilized.

TABLE 5. F-MEASURE FOR EACH EVENT TYPE (CLASS) AND EACH CLASSIFIER

| Classifier / Event type | J48 | Random Forest | Simple Cart | BayesNet | NaiveBayes | Decision Table | Ibk |
|---|---|---|---|---|---|---|---|
| merging | 0.97 | 0.97 | 0.97 | 0.83 | 0.87 | 0.96 | 0.92 |
| splitting | 1.00 | 1.00 | 1.00 | 1.00 | 1.00 | 1.00 | 1.00 |
| dissolving | 0.96 | 0.97 | 0.95 | 0.58 | 0.20 | 0.87 | 0.64 |
| continuing | 0.57 | 0.57 | 0.57 | 0.23 | 0.01 | 0.57 | 0.34 |
| shrinking | 0.67 | 0.59 | 0.68 | 0.42 | 0.09 | 0.64 | 0.50 |
| growing | 0.70 | 0.69 | 0.61 | 0.33 | 0.06 | 0.63 | 0.22 |

In the Table 5 and Figures 7-9 the F-measure comparison for all event types (classes) and all classifiers is presented. The three tree classifiers achieved the best results (the worst F-measure value is 0.57 for continuing), from the rest the Decision Table also achieved quite good results.

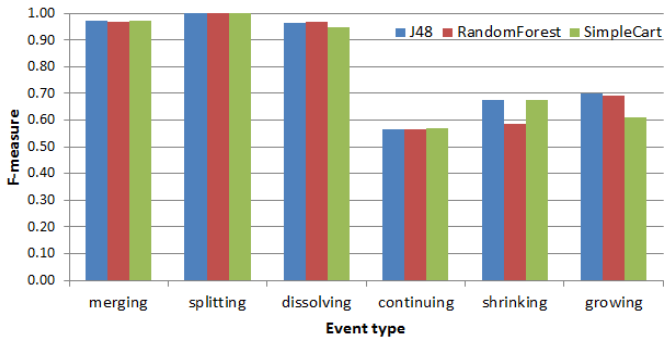

**Figure 7.** Results of event classification for decision tree classifiers

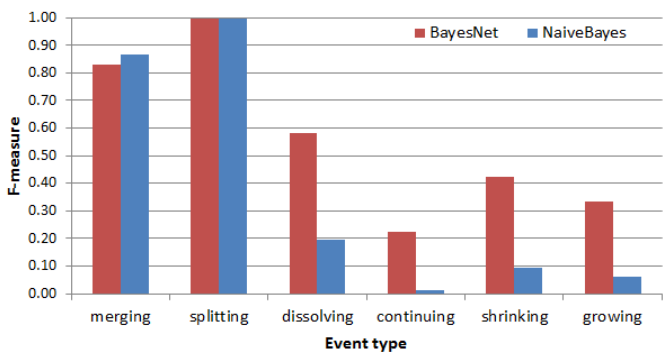

**Figure 8.** Results of event classification for probabilistic classifiers

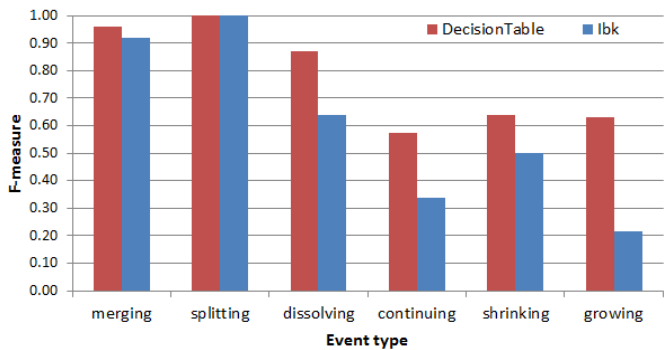

**Figure 9.** Results of event classification for other classifiers

As we can see each classifier achieves the best results for splitting, merging and dissolving events and the worst for continuing, shrinking and growing. This happens because of uneven distribution of different event types instances. This distribution was presented in Figure 10. The number of splitting events is much higher than for the rest of events. We think this is because the time frame size is too short for the most communities and they continuously splits and merge as service users migrates from one topic to another. Authors of the *GED* method suggests in [27] that increasing the size of the time frame increases the possibility for the emergence of persistent groups and this will be our next step in future work.

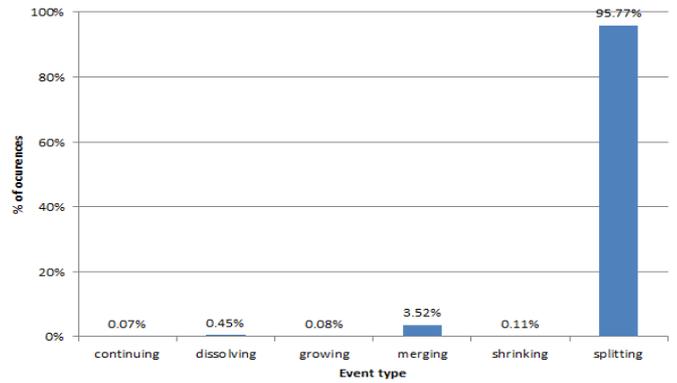

**Figure 10.** The percentage of each event cases in training dataset.

For the merging and dissolving events, most classifiers (for Naive Bayes, BayesNet and IBk the F-measure is lower) are able to produce very good results, despite the fact that they constitute only a small fraction of all events. Unfortunately, the number of continuing, shrinking and growing cases (around 0.1% of all cases) in training dataset is just too small to produce as impressive results as for other three events. However, the F-measure between 0.57 and 0.7 is a very good achievement especially for such uneven event distribution, once again Naive Bayes, BayesNet and IBk are exception and have lower F-measure. The poor results for Naive Bayes and BayesNet suggest that the probabilistic classifiers are not suited for predicting group evolution.

VI. DISCUSSION, CONCLUSIONS AND FUTURE WORK

In this paper the new method for future event prediction based on stable group changes identification algorithm (*SGCI*) has been presented together with the comparison to existing method, based on group evolution discovery algorithm (*GED*).

The methods differ in approach to community identification (fugitive or stable), event definition, kind of information used in classification (only attributes or attributes with previous events) and length of the considered community history.

In general, a high level of prediction quality was obtained using both presented methods. The best results, in the case of both methods, were obtained using different decision tree classifiers. *GED* method identifies disintegrations of group better while *SGCI* gives better results for change in size and simple group continuation without size changes. Merges and splits are similarly well identified in both methods. The worst prediction method results in terms of quality were obtained using Naive Bayesian classifier in the *SGCI* method and Naive Bayesian and Bayes Network classifiers in *GED*.

If one compares the total results achieved by both methods, one can notice that they are quite similar, however in some cases, using *SGCI* can improve the achieved results. This might be as a result of the *SGCI* design, i.e. that the method is already tailored to extract persistent communities, while the *GED* requires a longer time frame to achieve this.

Future research may be performed in few directions. The first being the analysis of recognizing attributes which have a higher influence on the prediction of given kinds of events. The second direction of research will be an analysis of how the length of the considered community histories influence the accuracy of the prediction. The third one will be an analysis of the event definitions and values of thresholds associated with the given events. In the fourth research direction, based on our results from this article, we are planning to use appropriate approaches to this problem as: undersampling or oversampling, cost sensitive learning (i.e. to change the cost function to punish more for the mistakes with underrepresented classes) and special algorithms and methods which can deal with unbalanced datasets e.g. SVM or AdaBoost. Next research direction will be to use different machine learning approaches to this problem e.g. SVM, logistics regression, AdaBoost, etc. Finally, we want to carry out experiments with unsupervised learning to better understand data used for prediction.

## REFERENCES

[1] D. Liben-Nowell, J. Kleinberg, "The link-prediction problem for social networks", J. Am. Soc. Inf. Sci., 58, 2007, pp 1019–1031, doi: 10.1002/asi.20591.

[2] E. Zheleva, L. Getoor, J. Golbeck, U. Kuter, "Using friendship ties and family circles for link prediction". SNAKDD'08, Springer-Verlag, 2008, pp 97-113.

[3] R. Lichtenwalter, J. T. Lussier, N. V. Chawla, "New perspectives and methods in link prediction", Proceedings of the 16th ACM SIGKDD International Conference on Knowledge Discovery and Data Mining, Washington, DC, USA, 25-28 July 2010, pp 243-252, ACM, New York.

[4] P. Symeonidis, E. Tiakas, Y. Manolopoulos, "Transitive node similarity for link prediction in social networks with positive and negative links", Proceedings of RecSys 2010, Barcelona, Spain, 26-30 September 2010, pp 183-190, ACM, New York.

[5] J. Leskovec, D. P. Huttenlocher, J. M. Kleinberg, "Predicting positive and negative links in online social networks", Proceedings of WWW 2010, Raleigh, North Carolina, USA, 26-30 April 2010, pp 641-650, ACM, New York.

[6] K. Y. Chiang, N. Natarajan, A. Tewari, I.S. Dhillon, "Exploiting longer cycles for link prediction in signed networks", Proceedings of CIKM 2011, Glasgow, United Kingdom, 24-28 October 2011, pp 1157-1162, ACM, New York.

[7] J. Kunegis, A. Lommatzsch, C. Bauckhage, "The slashdot zoo: mining a social network with negative edges", Proceedings of WWW 2009, Madrid, Spain, 20-24 April 2009, pp 741-750, ACM, New York.

[8] D. Davis, R. Lichtenwalter, N. V. Chawla, "Supervised methods for multi-relational link prediction", Social Network Analysis and Mining, Springer Vienna, 2012, doi: 10.1007/s13278-012-0068-6.

[9] Y. Richter, E. Yom-Tov, N. Slonim, "Predicting Customer Churn in Mobile Networks through Analysis of Social Groups", Proceedings of SDM 2010, Columbus, Ohio, USA, May 2010, pp 732-741, SIAM, Philadelphia, PA.

[10] A. Wai-Ho, K. C. C. Chan, Y. Xin, "A novel evolutionary data mining algorithm with applications to churn prediction", IEEE Transactions on Evolutionary Computation, 2003, vol. 7, no. 6, pp 532-545.

[11] B. Gliwa, S. Saganowski, A. Zygmunt, P. Bródka, P. Kazienko, J. Koźlak, "Identification of Group Changes in Blogosphere", ASONAM 2012, IEEE Computer Society, 2012.

[12] A. Zygmunt, P. Bródka, P. Kazienko, J. Koźlak, "Key person analysis in social communities within the blogosphere", J. UCS 18(4), pp. 577-597 (2012).

[13] B. Gliwa, A. Zygmunt, A. Byrski, "Graphical analysis of social group dynamics", CASoN 2012, IEEE Computer Society, 2012.

[14] P. Bródka, S. Saganowski, P. Kazienko, "*GED*: The Method for Group Evolution Discovery in Social Networks", Social Network Analysis and Mining, March 2013, Volume 3, Issue 1, pp 1-14.

[15] L.C. Freeman, "Centrality in Social Networks. Conceptual Clarification", Social Networks 1 (1978/79), pp. 215-239.

[16] S. Wasserman, K. Faust, "Social Network Analysis: Methods and Applications", Cambridge University Press, 1994.

[17] P. Bródka, P. Kazienko, B. Kołoszczyk, "Predicting Group Evolution in the Social Network", SocInfo 2012, LNCS 7710, Springer, 2012, pp. 54–67.

[18] P. Bródka, K. Musiał, P. Kazienko, "A Performance of Centrality Calculation in Social Networks", CASoN 2009, IEEE Computer Society, 2009, pp.24-31.

[19] L. Breiman, J.H. Friedman, R.A. Olshen, C.J. Stone, "Classification and regression trees", Monterey CA: Wadsworth & Brooks/Cole Advanced Books & Software 1984.

[20] R. Quinlan, "C4.5: Programs for Machine Learning", Morgan Kaufmann Publishers, San Mateo, CA, 1993.

[21] L. Breiman, "Random Forests", Machine Learning. vol. 45, issue 1, 2001, pp. 5-32.

[22] M. Hall, E. Frank, G. Holmes, B. Pfahringer, P. Reutemann, I.H. Witten, The WEKA Data Mining Software: An Update, SIGKDD Explorations, 2009, vol. 11, issue 1.

[23] G.H. John, P. Langley, "Estimating Continuous Distributions in Bayesian Classifiers", Eleventh Conference on Uncertainty in Artificial Intelligence, San Mateo, 1995, pp. 338-345.

[24] D. Aha, D. Kibler, "Instance-based learning algorithms". Machine Learning. vol. 6, 1991, pp. 37-66.

[25] R. Kohavi, "The Power of Decision Tables", 8th European Conference on Machine Learning, 1995, pp. 174-189.

[26] G.J. McLachlan, K.A. Do, C. Ambroise, "Analyzing Microarray Gene Expression Data, Wiley Series in Probability and Statistics", 2004, ISBN-10: 0471226165.

[27] S. Saganowski, P. Bródka, P. Kazienko, "Influence of the Dynamic Social Network Timeframe Type and Size on the Group Evolution Discovery", ASONAM 2012, IEEE Computer Society, 2012, pp. 678-682.